\documentclass[aps,prl,twocolumn,nofootinbib,groupedaddress,superscriptaddress
,longbibliography]{revtex4-2}

\usepackage{amsmath,amssymb}
\usepackage{graphicx}
\usepackage{multirow}
\usepackage{bm}
\usepackage{mathtools}
\usepackage{dsfont}
\usepackage{amsfonts}
\usepackage{xcolor}
\usepackage[normalem]{ulem}
\usepackage{url}
\usepackage{wasysym}
\usepackage{bbm}
\usepackage{hyperref}
\usepackage{braket}
\usepackage{booktabs}

\begin{document}

\title{Quantum Geometric Origin of Yu-Shiba-Rusinov States}

\author{Rui-Xing Zhang}
\email{ruixing@utk.edu}
\address{Department of Physics and Astronomy, The University of Tennessee, Knoxville, Tennessee 37996, USA}
\address{Department of Materials Science and Engineering, The University of Tennessee, Knoxville, Tennessee 37996, USA}

\begin{abstract}
In this work, we revisit the classic Yu-Shiba-Rusinov (YSR) problem of magnetic impurities in $s$-wave superconductors and uncover a quantum geometric origin of the YSR bound-state structure. The key geometric quantity is the momentum-space average of Bloch band projector over low-energy electrons, whose matrix rank, positive eigenvalues, and corresponding eigenvectors directly control the number, energies, and symmetry representations of the YSR states, respectively. As a result, a superconductor with nontrivial normal-state quantum geometry can host more bound states than its trivial counterpart, even if they share the same electronic dispersion. We apply our theory to monolayer $1H$-NbSe$_2$ and find that the geometric texture of its single superconducting band enforces three symmetry-distinct YSR channels for a magnetic point impurity at Nb or Se sites. Our work opens a route to probing normal-state wavefunction geometry through local YSR spectroscopy.
\end{abstract}

\maketitle

\section{Introduction} 
In the 1960s, Yu~\cite{yu1965bound}, Shiba~\cite{shiba1968}, and Rusinov~\cite{rusinov1969} independently showed that magnetic impurities in a conventional $s$-wave superconductor (SC) can locally break Cooper pairs and trap quasiparticle states inside the superconducting gap. Both energies and spatial profiles of these Yu--Shiba--Rusinov (YSR) states can be precisely characterized by modern scanning tunneling microscopy (STM) techniques~\cite{yazdani1997probing}, which carry important electronic and pairing information of the underlying SCs~\cite{hoffman2002imaging,mcelroy2003relating,hanaguri2010unconventional,kreisel2015qpi,ming2023evidence,wu2026prx}. For instance, spatial oscillations of YSR states reflect characteristic Fermi momenta~\cite{ruby2016orbital}, while directional focusing reflects the anisotropy of the normal-state Fermi surface~\cite{menard2015coherent,kim2020long,simon2022focusing}. More broadly, Fourier analysis of impurity-induced quasiparticle interference patterns has become a widely used approach to probing band dispersions and Fermi-surface structure~\cite{wang2003qpi,balatsky2006rmp}.

Bloch wavefunctions, however, carry information that the band dispersion alone cannot specify. For example, two bands with identical dispersions yet different wavefunction patterns can exhibit qualitatively distinct topological responses, such as their Hall conductance~\cite{thouless1982,haldane1988}. Even when both dispersion and topology are fixed, Bloch wavefunctions can vary differently across momentum space, as captured by the notion of quantum geometry~\cite{provost1980riemannian,torma2023essay,yu2025quantum,liu2025quantum}, which can influence observables such as nonlinear transport~\cite{sodemann2015dipole}, electron-phonon coupling~\cite{yu2024non}, and the superfluid stiffness of SCs~\cite{peotta2015superfluidity}.  However, whether and how electron wavefunctions contribute to the YSR physics remains largely unexplored. 

In this work, we show that Bloch states quantitatively control the number, energies, and wavefunctions of YSR states through a normal-state quantum-geometric quantity, the {\it averaged band projector}. We prove this finding analytically for $s$-wave SCs hosting a point-like, classical magnetic impurity with orbital-independent exchange coupling. For a flat-band SC, this quantity is the Brillouin-zone (BZ) average $Q=N^{-1}\sum_{\bf k}P({\bf k})$, where $P({\bf k})$ is the Bloch-state projector of the target band. For a dispersive band in the weak-pairing limit, $Q$ is replaced by its density-of-states-weighted Fermi-surface counterpart $Q_{\rm FS}$. Specifically, the positive eigenvalues of $Q$ or $Q_{\rm FS}$ determine the YSR energies, while the corresponding eigenvectors determine the orbital character of the bound states and hence their symmetry representations. 

In the textbook single-band YSR model~\cite{yu1965bound,shiba1968,rusinov1969}, a point magnetic impurity produces a single particle-hole-related pair of bound states. This classic result takes the band dispersion as the normal-state input, but not the corresponding Bloch wavefunction. Our quantum geometric theory reveals a more general counting principle: the number of YSR pairs is dictated by the {\it matrix rank} of the corresponding $Q$ or $Q_{\rm FS}$, rather than by band counting alone. Consequently, even a single superconducting band can harbor multiple YSR pairs purely enabled by its inherent quantum geometry. We verify this prediction in a minimal flat-band SC model, finding quantitative agreement between analytical and numerical results for both the bound-state energies and wavefunctions.

We further propose monolayer $1H$-NbSe$_2$~\cite{ugeda2016characterization,xi2016ising} as a realistic platform for exploring quantum-geometry-enabled YSR states. As a 2D single-band SC, the superconducting band of $1H$-NbSe$_2$ has recently been identified as an obstructed atomic band with nontrivial quantum geometry~\cite{yu2026NbSe2,cualuguaru2026observation,holbrook2026real}. Using a realistic 12-band model, we evaluate $Q_{\rm FS}$ and find that this single band supports three YSR channels distinguished by their $C_3$ eigenvalues for magnetic impurities on either Nb or Se sites. Full $T$-matrix and real-space BdG calculations quantitatively reproduce the predicted YSR energies and confirm their symmetry character. We also predict a pronounced contrast between the YSR spectra of Nb- and Se-site impurities, which can be directly tested in experiments.

\section{Band-projected $T$-matrix theory}
We consider a time-reversal-invariant SC with conventional, orbital-independent $s$-wave spin-singlet pairing. For simplicity, we assume the normal state to have vanishing Rashba spin-orbit coupling. The normal-state Hamiltonian is $h({\bf k})=h_\uparrow({\bf k})\oplus h_\downarrow({\bf k})$ with $h_\uparrow({\bf k}) = h^*_\downarrow(-{\bf k})$, as required by time-reversal symmetry (TRS). The basis vector for each spin sector is $c_{\mathbf{k},s}=(c_{\mathbf{k},1,s},c_{\mathbf{k},2,s},\ldots,c_{\mathbf{k},M,s})^T$, where $c_{\mathbf{k},\alpha,s}$ annihilates an electron in orbital $\alpha=1,\ldots,M$ with spin index $s\in\{\uparrow,\downarrow\}$. In the mean-field theory, the corresponding Bogoliubov-de Gennes (BdG) Hamiltonian matrix is block diagonal ${\cal H}({\bf k}) = H_\uparrow({\bf k})\oplus H_\downarrow({\bf k})$, where
\begin{equation}
    H_s({\bf k}) = \tau_z\otimes[h_s({\bf k})-\mu I_M] + \Delta \tau_x \otimes I_M,
\end{equation}
under the Nambu spinors $\Phi_{\mathbf{k},+}=(c_{\mathbf{k},\uparrow}^T,c_{-\mathbf{k},\downarrow}^{\dagger})^T$ and $\Phi_{\mathbf{k},-}=(c_{\mathbf{k},\downarrow}^T,-c_{-\mathbf{k},\uparrow}^{\dagger})^T$. Here, $\mu$ is the chemical potential, $\tau_{x,y,z}$ denote Pauli matrices for the particle-hole degree of freedom, $I_M$ is the $M\times M$ identity matrix in the orbital space, and we have picked a gauge choice in which $\Delta>0$. We further couple the SC to a point magnetic impurity at ${\bf r}=0$, with its moment along the $z$ direction and an {\it orbital-independent} exchange coupling strength $J$. The impurity potentials are thus $V_{\uparrow/\downarrow}=\pm J \tau_0 \otimes I_M$ in the spin-up and spin-down sectors, respectively. 

The YSR bound-state physics is encoded in the $T$ matrix, $T_s(E) = \left[V_s^{-1}-{\cal G}_s(E)\right]^{-1}$, where ${\cal G}_s(E)$ is the on-site, spin-resolved retarded Green's function for the clean SC. Specifically, ${\cal G}_s(E)=1/N\sum_{\bf k} G_s(E,{\bf k})$ is an average of the $k$-space Green's function $G_s(E,{\bf k}) = \left[E+i0^+-H_s({\bf k})\right]^{-1}$, with $N$ the number of unit cells. A subgap YSR state appears whenever $T_s(E)$ has a pole. Since particle-hole symmetry (PHS) maps a pole at energy $E$ in the spin-up sector to one at $-E$ in the spin-down sector, we focus on the spin-up sector and suppress the spin index $s$ unless otherwise stated, with which we denote $h({\bf k})\equiv h_\uparrow({\bf k})$.

We now rewrite the $T$ matrix with the normal-state bands by introducing a unitary matrix ${\cal U}_{\bf k}=\tau_0\otimes U_{\bf k}$. Here, $U_{\bf k}=(|u_{1}({\bf k})\rangle,\ldots,|u_{M}({\bf k})\rangle)$ consists of the normal-state eigenvectors with $h({\bf k})|u_{m}({\bf k})\rangle = \varepsilon_{m}({\bf k})|u_{m}({\bf k})\rangle$. Clearly, $U_{\bf k}$ diagonalizes the normal-state Hamiltonian with $U_{\bf k}^\dagger[h({\bf k})-\mu]U_{\bf k}=\text{diag}[\xi_{1}({\bf k}),\ldots,\xi_{M}({\bf k})]$, where we have defined $\xi_{m}({\bf k})\equiv \varepsilon_m({\bf k})-\mu$. Since the pairing term is proportional to $I_M$, there is no interband pairing after the ${\cal U}$ transformation. Consequently, $g(E,{\bf k})$, the free Green's function in the band basis, naturally separates into a set of decoupled $2\times 2$ blocks $g_{m}(E,{\bf k})=[(E+i0^+)\tau_0-\xi_{m}({\bf k})\tau_z-\Delta\tau_x]^{-1}$. Going back to the orbital basis, the on-site Green's function is thus
\begin{equation}
    {\cal G}(E) = \sum_m[\frac{1}{N}\sum_{\bf k}g_m\left(E,{\bf k}\right) \otimes P_{m}({\bf k})],
    \label{eq:projected Green function}
\end{equation}
where $P_{m}({\bf k})=|u_{m}({\bf k})\rangle\langle u_{m}({\bf k})|$ is the projector for the $m$-th Bloch band. Note that Eq.~\ref{eq:projected Green function} is exact. 

We next consider an {\it isolated-band limit}, where the target band $n$ is the only low-energy band around the Fermi level and participates in the superconductivity. We hence expect the remote-band contribution of ${\cal G}(E)$ with $m\neq n$ to be small. As a result, ${\cal G}(E)\approx {\cal G}_n(E) = \frac{1}{N}\sum_{\bf k}g_n\left(E,{\bf k}\right) \otimes P_{n}({\bf k})$ and the $T$ matrix is given by
\begin{equation}
    T^{-1}(E) = V^{-1} - {\cal G}_n(E).  
\label{eq:target_T_matrix}
\end{equation}

\section{Flat-band limit and averaged projector} 

We first consider a flat-band limit where the target band is dispersionless across the BZ, with $\xi_n({\bf k})=\xi_0$. Since the band Green's function $g_n(E,{\bf k}) = g_n(E)$ is now ${\bf k}$-independent, the on-site Green's function factorizes as ${\cal G}(E)=g_n(E) \otimes Q$. Here we have defined the {\it BZ-averaged projector},
\begin{equation}
    Q=\frac{1}{N}\sum_{\bf k}P_n({\bf k}).
    \label{eq:BZ_averaged_projector}
\end{equation}
Although $Q$ is defined as an average of projectors, it is generally not a projector itself. As proved in the Supplemental Material (SM)~\cite{supp}, $Q^2=Q$ if and only if $P_n({\bf k})$ is momentum independent, which corresponds to trivial quantum geometry. By construction, $Q$ is Hermitian and positive semidefinite, with $\operatorname{Tr}Q=1$. We denote its nonzero eigenvalues and the corresponding normalized eigenvectors by
\begin{equation}
    Q|\phi_a\rangle=\lambda_a|\phi_a\rangle,
    \qquad a=1,\ldots,\operatorname{rank}Q,
    \label{eq:Q_eigensystem}
\end{equation}
where $1\leq \operatorname{rank}Q \leq M$. The eigenvalues obey $0<\lambda_a\leq1$ and $\sum_{a}\lambda_a=1$. It is crucial to note that trivial quantum geometry leads to $\operatorname{rank}Q=1$, whereas nontrivial geometry necessarily features $Q^2\neq Q$ and $\operatorname{rank}Q>1$. In the following, we will prove that the eigensystem of $Q$ {\it fully determines} the YSR states of the flat-band SC.

At a pole energy $E$, $T^{-1}(E)$ has a zero eigenvalue, and we denote the corresponding eigenvector by $\chi$, with $T^{-1}(E)\chi=0$. As shown in the SM~\cite{supp}, $\chi=V\Psi(0)$, where $\Psi(0)$ is the YSR wavefunction at the impurity site. Since $g_n(E)$ and $Q$ act in the Nambu and orbital spaces, respectively, the corresponding eigenvectors can be chosen in the product form
\begin{equation}
    \chi_{\nu a}\propto|\eta_\nu\rangle\otimes|\phi_a\rangle,
\end{equation}
up to an overall normalization factor. Here, the normalized Nambu spinor $|\eta_\nu\rangle$ diagonalizes $g_n(E)$ with an eigenvalue of $(E-\nu{\cal E}_0)^{-1}$, where $\nu=\pm1$ and ${\cal E}_0=\sqrt{\xi_0^2+\Delta^2}$. 

For the spin-$\uparrow$ sector, $V_\uparrow=J\tau_0\otimes I_M$. By solving the pole equation, we find $E_{\uparrow\nu a}=\nu{\cal E}_0+J\lambda_a$, where we have restored the spin index for clarity. For the spin-$\downarrow$ sector, TRS requires the corresponding averaged projector to be $Q^*$, so that $Q^*|\phi_a^*\rangle=\lambda_a|\phi_a^*\rangle$. Noting that $V_\downarrow=-J\tau_0\otimes I_M$, we find $E_{\downarrow\nu a}=\nu{\cal E}_0-J\lambda_a$. Taken together, the bound-state energies are given by
\begin{equation}
    E_{\uparrow/\downarrow,\nu a}=\nu {\cal E}_0 \pm J\lambda_a,
    \label{eq:flat-band-energies}
\end{equation}
where the PHS flips both $s$ and $\nu$ indices. Note that the BdG gap is defined by $|E|<{\cal E}_0$. Considering $0<J<2{\cal E}_0/\lambda_a$, each nonzero $\lambda_a$ branch thus generates one pair of {\it in-gap} YSR states $\{E_{\uparrow-a}, E_{\downarrow+a}\}$. Since $\lambda_a\leq 1$, for $|J|<2{\cal E}_0$, the total number of in-gap YSR pairs $N_{\mathrm{YSR}}$ thus equals the number of nonzero eigenvalues of the averaged projector $Q$, with
\begin{equation}
    N_{\mathrm{YSR}}=\operatorname{rank}Q,
    \label{eq:flat_band_YSR_counting}
\end{equation}
where degeneracies are counted. Eq.~\ref{eq:flat_band_YSR_counting} is one of the key results of this work. 

The full real-space YSR wavefunction can now be reconstructed from $\chi$, or equivalently $\Psi(0)$. We find that the normalized wavefunction is~\cite{supp}
\begin{equation}
    \Psi_{\uparrow\nu a}({\bf r})
    =
    |\eta_\nu\rangle\otimes
    \frac{{\cal Q}({\bf r})|\phi_a\rangle}{\sqrt{\lambda_a}},
    \label{eq:flat_band_wavefunction}
\end{equation}
where ${\cal Q}({\bf r})=N^{-1}\sum_{\bf k}e^{i{\bf k}\cdot{\bf r}}P_n({\bf k})$ is a quantum geometric quantity dubbed the {\it real-space band projector}. Specifically, ${\cal Q}({\bf r})|\phi_a\rangle$ gives the spatial orbital profile by projecting the impurity-site orbital state $|\phi_a\rangle$ onto the target flat band. At the impurity site, ${\cal Q}(0)=Q$ and $\Psi_{\uparrow\nu a}(0)=\sqrt{\lambda_a}|\eta_\nu\rangle\otimes|\phi_a\rangle = J^{-1}\chi_{\nu a}$. Since both $|\eta_\nu\rangle$ and $|\phi_a\rangle$ are normalized, 
\begin{equation}
    |\Psi_{\uparrow \nu a}(0)|^2 = \lambda_a.
\end{equation}
Thus, the value of $\lambda_a$ is exactly the impurity-site weight of the YSR state. The corresponding spin-$\downarrow$ partner can be constructed similarly with a conjugate orbital profile ${\cal Q}^*({\bf r})|\phi_a^*\rangle/\sqrt{\lambda_a}$.

We now explore the symmetry of $\Psi_{\uparrow\nu a}({\bf r})$. Let $A$ be a unitary point-group symmetry of the SC that is preserved by the magnetic impurity, and we use $\widehat{A}$ to denote its action on the orbital degrees of freedom. The target-band projector is covariant under $A$, with $P_n(A{\bf k})=\widehat{A}P_n({\bf k})\widehat{A}^\dagger$. We immediately find that
\begin{equation}
    {\cal Q}(A{\bf r})
    =
    \widehat{A}{\cal Q}({\bf r})\widehat{A}^\dagger,
    \qquad
    [Q,\widehat{A}]=0.
    \label{eq:Q_symmetry}
\end{equation} 
The eigenvectors of $Q$ can hence be chosen to transform under irreducible representations (irreps) of the impurity-site symmetry group. Specifically, for an irrep $\alpha$, $\widehat{A}|\phi_a\rangle=\sum_b|\phi_b\rangle D_{ba}^{(\alpha)}(A)$, where $D^{(\alpha)}(A)$ is the corresponding representation matrix and the sum runs over states within the same irrep multiplet. Furthermore, $\widehat{A}$ acts identically on the Nambu spinor $|\eta_\nu\rangle$. Because $S_z$ is conserved, the physical spin contributes only a fixed phase within each spin sector, which can be included when assigning the full double-group representations. Following Eq.~\ref{eq:flat_band_wavefunction}, we thus arrive at
\begin{equation}
    \widehat{A}\Psi_{\uparrow\nu a}({\bf r})
    =
    \sum_b
    \Psi_{\uparrow\nu b}(A{\bf r})D_{ba}^{(\alpha)}(A),
    \label{eq:YSR_symmetry}
\end{equation}
where the YSR wavefunctions are found to transform under {\it the same irreps} as the corresponding $|\phi_a\rangle$. The spin-$\downarrow$ particle-hole partner transforms under the complex-conjugate irrep. Therefore, the eigenvalues and eigenvectors of $Q$ govern the energies and symmetry characters of the YSR states in a flat-band SC, respectively.

\section{Minimal two-band flat-band model} 

As a proof of concept, we now present a 1D spin-degenerate flat-band SC model, whose YSR physics will be explored both analytically and numerically. For the spin-$\uparrow$ sector, the normal-state Hamiltonian is $h_\theta(k)={\bf n}_\theta(k)\cdot{\boldsymbol\sigma}$, where $\theta\in[0,\pi]$ is a tunable parameter and
\begin{equation}
    {\bf n}_\theta(k)
    =
    \left(
        \sin\theta\cos k,\,
        \sin\theta\sin k,\,
        \cos\theta
    \right).
    \label{eq:two_band_model}
\end{equation}
Here, ${\boldsymbol\sigma}=(\sigma_x,\sigma_y,\sigma_z)$ denotes the Pauli matrices acting on the two orbitals. Since $h_\theta^*(-k)=h_\theta(k)$, TRS gives $h_\downarrow(k)=h_\theta(k)$ for the spin-$\downarrow$ sector. Since $|{\bf n}_\theta(k)|=1$, the two bands are hence exactly flat at $\varepsilon_\pm=\pm1$ for every $\theta$. We choose the lower band as the target band, whose energy relative to the chemical potential is $\xi_0=-1-\mu$. 

As $k$ traverses the BZ, ${\bf n}_\theta(k)$ traces a circle at the polar angle $\theta$ on the Bloch sphere. At $\theta=0$, this circle collapses to the north pole and $h_0(k)=\sigma_z$ describes a $k$-independent trivial atomic limit, where the lower-band Wannier function is strictly localized on the $\sigma_z=-1$ orbital. At $\theta=\pi/2$, ${\bf n}_{\pi/2}(k)$ winds once around the equator and $h_{\pi/2}(k)=\cos k\,\sigma_x+\sin k\,\sigma_y$ exactly reproduces the fully dimerized topological limit of the Su-Schrieffer-Heeger model. In this limit, the Wannier function is localized on an intercell bond. At $\theta=\pi$, the model reaches the complementary atomic limit $h_\pi(k)=-\sigma_z$. This continuous evolution of the Wannier center is captured by the $\theta$-dependence of electric polarization $\gamma(\theta)=(1-\cos\theta)/2$, which achieves a ``Berry dipole" in the $(\theta,k)$ space~\cite{zhu2023scattering}. Therefore, varying $\theta$ effectively changes the Bloch-wavefunction texture of the system, while leaving the band dispersion invariant.

We next apply the flat-band results to $h_\theta(k)$ with $s$-wave spin-singlet pairing $\Delta$ and a point magnetic impurity with an exchange coupling strength of $J$. The projector onto the target band is $P(k)=[I_2-h_\theta(k)]/2$. Its BZ average is naturally diagonal $Q=\text{diag}(\lambda_1,\lambda_2)$, with eigenvalues $\lambda_1=\sin^2(\theta/2)$ and $\lambda_2=\cos^2(\theta/2)$. We choose the corresponding normalized eigenvectors as $|\phi_1\rangle=(1,0)^T$ and $|\phi_2\rangle=(0,1)^T$. Following Eq.~\ref{eq:flat-band-energies}, the in-gap YSR bound-state energies are 
\begin{eqnarray}
    E_{\uparrow-1}&=&-{\cal E}_0+J\sin^2(\theta/2),\nonumber \\
    E_{\uparrow-2}&=&-{\cal E}_0+J\cos^2(\theta/2),
    \label{eq:two-band energy}
\end{eqnarray}
as well as their spin-$\downarrow$ counterparts with opposite energies, where ${\cal E}_0=\sqrt{(\mu+1)^2+\Delta^2}$. Notably, at $\theta=0$ or $\pi$, $P(k)$ is momentum independent and the band quantum geometry is trivial. One eigenvalue of $Q$ vanishes, while the remaining eigenvalue produces a single in-gap YSR pair.
For $0<\theta<\pi$, both eigenvalues are nonzero with $\operatorname{rank}Q=2$, and two in-gap YSR pairs appear. In particular, the two YSR pairs become degenerate at $\theta=\pi/2$. Since the electronic spectrum is independent of $\theta$, the evolution of the YSR states arises entirely from the variation of the normal-state quantum geometry.

\begin{figure}[t]
\includegraphics[width=0.48\textwidth]{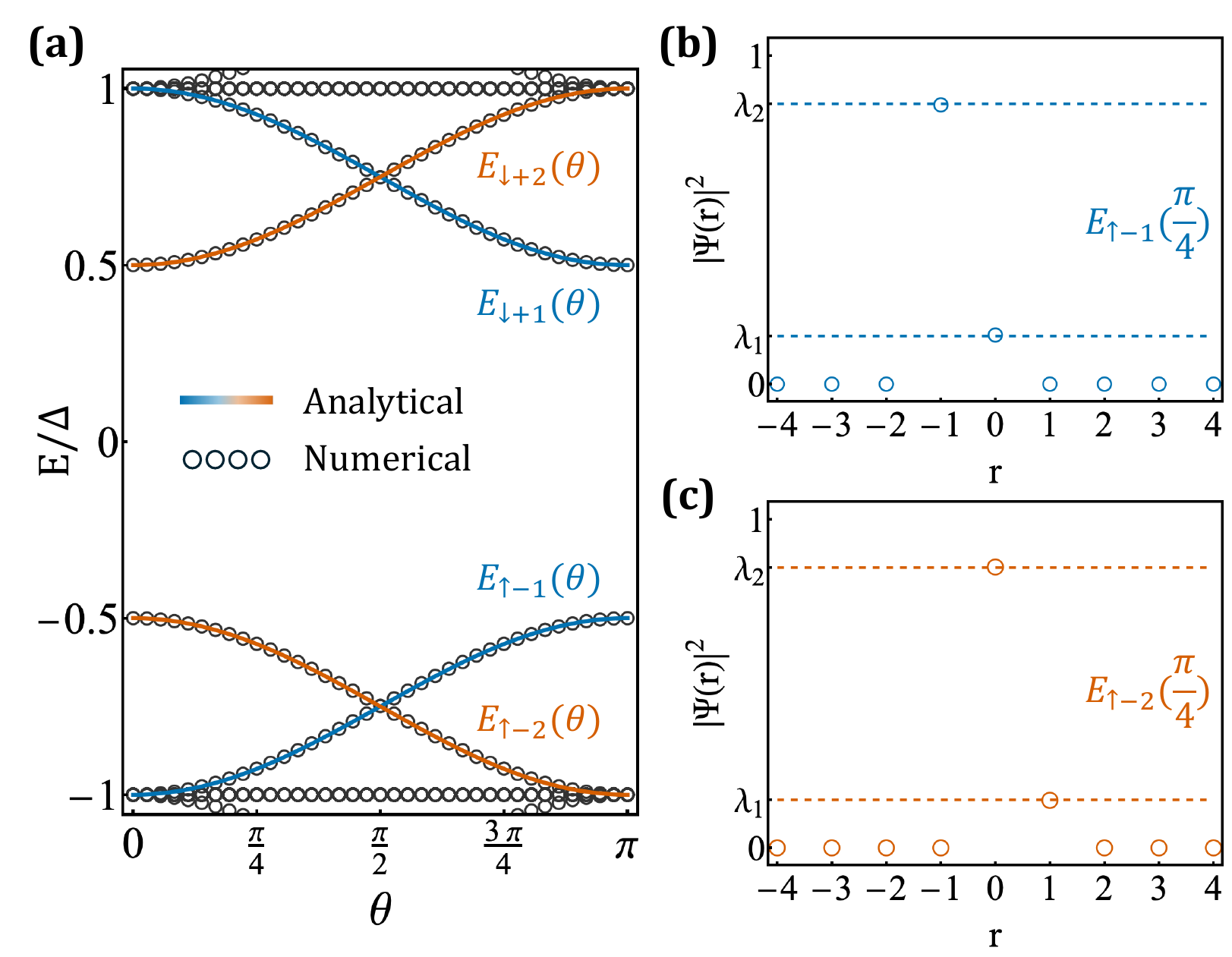}
    \caption{YSR states in the two-band flat-band model. (a) BdG spectrum as a function of $\theta$, where the impurity-induced YSR states live at $|E|<\Delta$. Open circles show exact-diagonalization results, and blue and red solid curves show the analytical YSR energies from the averaged projector $Q$. (b,c) Unit-cell probability densities of $\Psi_{\uparrow-1}$ and $\Psi_{\uparrow-2}$ at $\theta=\pi/4$, respectively. The impurity sits at $r=0$. Open circles show the numerical results and dashed lines mark the analytical weights $\lambda_1=\sin^2(\pi/8)$ and $\lambda_2=\cos^2(\pi/8)$. }
    \label{fig:flat band}
\end{figure}

Since $P(k)$ only involves constant and $e^{\pm ik}$ harmonics, its Fourier transform gives the real-space projector, 
\begin{equation}
    {\cal Q}(r) = \frac{1}{2}\begin{pmatrix}
        (1-\cos\theta)\delta_{r,0} & -\sin\theta \delta_{r,1} \\
        -\sin\theta \delta_{r,-1} & (1+\cos\theta)\delta_{r,0}\\
    \end{pmatrix},
\end{equation}
where $r$ is a unit-cell label relative to the impurity and $\delta_{r,r'}$ is the Kronecker delta. Following Eq.~\ref{eq:flat_band_wavefunction}, we arrive at the YSR wavefunctions by applying ${\cal Q}(r)$ to $|\phi_{1,2}\rangle$:
\begin{equation}
\begin{aligned}
    \Psi_{\uparrow-1}(r)
    &=
    |\eta_-\rangle\otimes
    \begin{pmatrix}
        \sin(\theta/2)\delta_{r,0}\\
        -\cos(\theta/2)\delta_{r,-1}
    \end{pmatrix},\\
    \Psi_{\uparrow-2}(r)
    &=
    |\eta_-\rangle\otimes
    \begin{pmatrix}
        -\sin(\theta/2)\delta_{r,1}\\
        \cos(\theta/2)\delta_{r,0}
    \end{pmatrix},
\end{aligned}
    \label{eq:two_band_YSR_wavefunctions}
\end{equation}
where the Nambu spinor $|\eta_-\rangle = [2{\cal E}_0({\cal E}_0+\xi_0)]^{-1/2}(-\Delta, {\cal E}_0+\xi_0)^T$. At $\theta=0$ or $\pi$, only the state associated with the nonzero eigenvalue remains an in-gap YSR state, and it is confined entirely to the impurity unit cell.
For $\theta\in(0,\pi)$, both YSR states are compactly localized over the impurity unit cell $r=0$ and one of its neighboring cells, i.e., $r=-1$ or $r=1$, with impurity-site weights $\sin^2(\theta/2)=\lambda_1$ and $\cos^2(\theta/2)=\lambda_2$, respectively. The spin-$\downarrow$ partners share the same spatial probability distributions due to the PHS. 

We next numerically verify the above predictions. We place this flat-band SC model on a periodic chain with $L=50$ unit cells, eliminating possible boundary states. Our choice of model parameters is $\mu=-1$ and $\Delta=0.2$, with which the Fermi level exactly crosses the lower flat band with $\xi_0=0$ and ${\cal E}_0=\Delta$. We further include a magnetic point impurity with $J=0.1$ at the origin $r=0$. The YSR spectrum as a function of $\theta$ is shown in Fig.~\ref{fig:flat band} (a), where the open circles denote the numerical spectrum via exact matrix diagonalization. Just as expected, for $0<\theta<\pi$, two YSR pairs appear inside the gap and become degenerate at $\theta=\pi/2$.  At $\theta=0$ and $\pi$, one particle-hole-related pair merges with the bulk BdG levels at $E=\pm\Delta$, leaving only one isolated in-gap YSR pair. The blue and orange solid curves represent the analytical YSR energies in Eq.~\ref{eq:two-band energy}, which quantitatively reproduce the numerical YSR spectra. Figs.~\ref{fig:flat band} (b) and (c) further compare the analytical and numerical spatial probability distributions for the two spin-$\uparrow$ YSR states at $\theta=\pi/4$, where $|\Psi(r)|^2$ is summed over the Nambu and orbital components. For $\Psi_{\uparrow-1}$, the probability weights at $r=0$ and $r=-1$ are found to be $\lambda_1=\sin^2(\pi/8)$ and $\lambda_2=\cos^2(\pi/8)$, respectively, whereas those of $\Psi_{\uparrow-2}$ at $r=0$ and $r=1$ are $\lambda_2$ and $\lambda_1$. The probability vanishes in all other unit cells, in perfect agreement with the analytical YSR wavefunctions in Eq.~\ref{eq:two_band_YSR_wavefunctions}. These results confirm the exactness of our quantum-geometric theory of YSR physics in the flat-band limit.

\section{Weak-pairing limit and Fermi-surface-averaged projector}

We now move away from the flat-band limit and proceed to explore YSR physics in a dispersive target band in the weak-pairing limit, where superconductivity is dominated by electrons near the Fermi surface (FS). Back to the form of ${\cal G}_n(E)$, we note that $g_n(E,{\bf k})$ becomes ${\bf k}$-dependent and can no longer be taken outside the ${\bf k}$-sum. Therefore, we rewrite the on-site Green's function as
\begin{equation}
    {\cal G}_n(E)
    =
    \int d\xi\,g_n(E,\xi)\otimes\rho(\xi),
    \label{eq:spectral_density}
\end{equation}
where $g_n(E,\xi)=[(E+i0^+)\tau_0-\xi\tau_z-\Delta\tau_x]^{-1}$ is obtained from $g_n(E,{\bf k})$ by replacing $\xi_n({\bf k})$ with $\xi$. Here, $\rho(\xi)$ is the matrix-valued spectral density defined by
\begin{equation}
    \rho(\xi)
    =
    \frac{1}{N}\sum_{\bf k}
    \delta\!\left[\xi-\xi_n({\bf k})\right]P_n({\bf k}),
    \label{eq:rho_BCS}
\end{equation}
where $\delta$ denotes the Dirac delta function. The matrix $\rho(\xi)$ describes the orbital-resolved spectral weight on the constant-energy surface $\xi_n({\bf k})=\xi$, while its trace gives the normal-state density of states (DOS) per spin. 

In the weak-pairing limit, we assume that $\rho(\xi)$ varies weakly within an energy window $|\xi|<\Lambda$ around the Fermi energy, where $\Delta\ll\Lambda$. We can hence approximate $\rho(\xi)\simeq\rho(0)$, which is the spectral density matrix at the Fermi level. This further leads to ${\cal G}_n(E)\simeq[\int_{-\Lambda}^{\Lambda}d\xi\,g_n(E,\xi)]\otimes\rho(0)$. This expression shares the same factorized structure as the flat-band result, with $\rho(0)$ replacing $Q$. However, unlike $Q$ with a unity trace, $\nu_F=\operatorname{Tr}\rho(0)$ is the normal-state DOS per spin at the Fermi level. This inspires us to define a ``normalized" band geometric quantity $Q_{\mathrm{FS}}=\rho(0)/\nu_F$, which is referred to as the {\it FS-averaged projector}. Based on Eq.~\ref{eq:rho_BCS}, we arrive at
\begin{equation}
    Q_{\mathrm{FS}}
    =
    \frac{1}{\nu_F}
    \int_{\mathrm{FS}}
    \frac{dS_{\bf k}}
    {(2\pi)^d\left|\nabla_{\bf k}\xi_n({\bf k})\right|}
    P_n({\bf k}).
\label{eq:FS_averaged_projector}
\end{equation}
Here, $d$ is the spatial dimension and $dS_{\bf k}$ is the surface element on the FS. We assume that the FS lies away from Lifshitz transitions, with
$\left|\nabla_{\bf k}\xi_n({\bf k})\right|\neq0$
everywhere on the FS.

Physically, $Q_{\mathrm{FS}}$ is the normalized, DOS-weighted FS average of the band projector. Just like its BZ-averaged counterpart, $Q_{\mathrm{FS}}$ is Hermitian and positive semidefinite with $\operatorname{Tr}Q_{\mathrm{FS}}=1$. Notably, $\operatorname{rank}Q_\mathrm{FS}>1$ as long as $P_n({\bf k})$ is not constant along the FS. The remaining $\xi$ integral of ${\cal G}_n$ can be evaluated analytically. As shown in the SM~\cite{supp}, for $|E|<\Delta$, we arrive at
\begin{equation}
    {\cal G}_n(E)
    \simeq
    -\pi\nu_F
    \frac{E\tau_0+\Delta\tau_x}
    {\sqrt{\Delta^2-E^2}}
    \otimes Q_{\mathrm{FS}},
    \label{eq:BCS_Green_function}
\end{equation}
which shares the same tensor-product structure as the flat-band Green's function. 

We denote the nonzero eigenvalues and corresponding normalized eigenvectors of $Q_{\mathrm{FS}}$ by $\lambda_a$ and $|\phi_a\rangle$, respectively, with $Q_{\mathrm{FS}}|\phi_a\rangle=\lambda_a|\phi_a\rangle$ and $a=1,\ldots,\operatorname{rank}Q_{\mathrm{FS}}$. Similar to the flat-band case, the eigenbasis of $Q_{\mathrm{FS}}$ decomposes $T^{-1}(E)$ into independent $2\times2$ Nambu blocks. Solving the pole equation of the corresponding $T$-matrix gives the in-gap YSR energies,
\begin{equation}
    E_{\uparrow/\downarrow,a}
    =
    \mp\Delta
    \frac{1-\left(\pi\nu_FJ\lambda_a\right)^2}
    {1+\left(\pi\nu_FJ\lambda_a\right)^2},
    \label{eq:BCS_YSR_energy}
\end{equation}
where we have restored the spin index for clarity. Physically, the FS quantum geometry encoded by $\{\lambda_a\}$ converts the exchange coupling strength $J$ into a set of effective couplings $J\lambda_a$. Therefore, within the weak-pairing limit, the number of YSR pairs is,
\begin{equation}
    N_{\rm YSR}=\operatorname{rank}Q_{\rm FS},
\end{equation}
with degeneracies counted. When $\operatorname{rank}Q_{\mathrm{FS}}=1$, the only nonzero eigenvalue is $\lambda_1=1$, and Eq.~\ref{eq:BCS_YSR_energy} reduces to the well-known conventional YSR solution in Refs.~\cite{yu1965bound,shiba1968,rusinov1969}. 

Similar to the flat-band case, for any unitary point-group symmetry $A$ preserved by the impurity, the covariance of $P_n({\bf k})$ and the invariance of the FS integration measure imply $[Q_{\mathrm{FS}},\widehat{A}]=0$. The eigenvectors $|\phi_a\rangle$ therefore carry irreps of the impurity-site group, which are further inherited by the corresponding YSR states. Thus, the full YSR information is now encoded in the eigensystem of $Q_{\mathrm{FS}}$ in the weak-pairing limit.

\section{Application to monolayer $\text{NbSe}_2$}

We now apply our theory to monolayer $1H$-NbSe$_2$. At low energies, a single time-reversal-related pair of Nb-derived bands forms the Fermi surface and supports intrinsic superconductivity with $T_c\sim3$K~\cite{ugeda2016characterization,xi2016ising}. Recently, however, the relevant band was shown to feature atomic obstruction, with its Wannier center lying away from the atomic sites~\cite{cualuguaru2026observation,yu2026NbSe2,holbrook2026real}. Although a simple single-orbital model can accurately reproduce the band dispersion of NbSe$_2$, it necessarily misses the multiorbital nature of the Bloch wavefunctions. This intrinsic quantum-geometric character thus makes monolayer $1H$-NbSe$_2$ a natural material testbed for our YSR theory.

\begin{figure*}[t]
\includegraphics[width=0.9\textwidth]{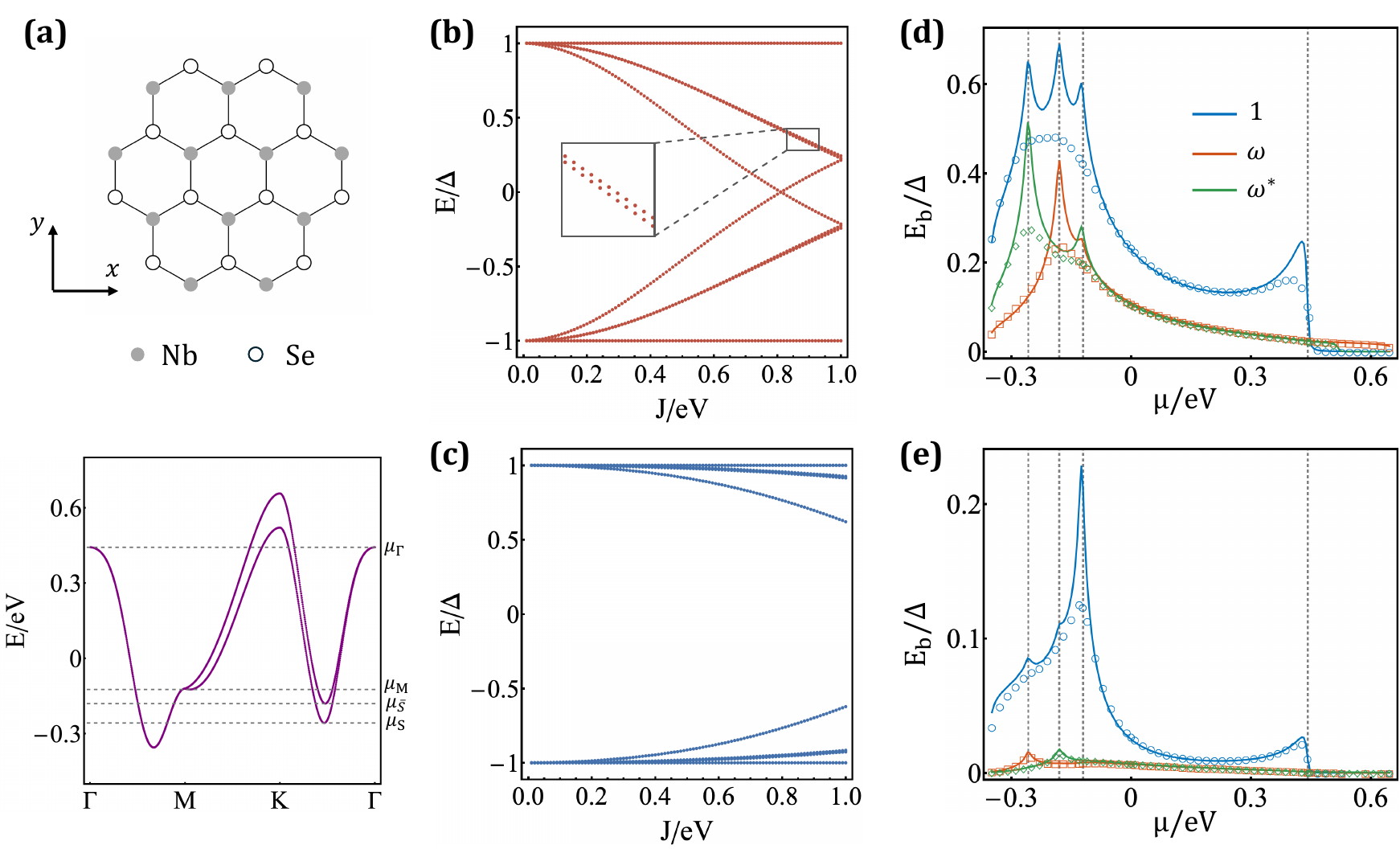}
    \caption{YSR characterization of monolayer $1H$-NbSe$_2$. (a) Lattice structure (top) and the low-energy band of interest calculated from the six-orbital tight-binding model (bottom). Dashed lines in (a), (d), and (e) mark the normal-state Lifshitz transitions. (b,c) YSR energies versus exchange strength $J$ at $\mu=0$ via exact diagonalization for Nb- and Se-centered impurities, respectively. The inset in (b) enlarges the $\omega$ and $\omega^*$ branches and resolves the small Ising-SOC-induced splitting. (d,e) YSR binding energies versus $\mu$ for Nb and Se impurities at $J=0.30\,\mathrm{eV}$. Solid curves denote the $Q_{\rm FS}^{\zeta}$-based predictions, while open symbols denote the full six-orbital $T$-matrix results. Colors distinguish the orbital $C_3$ channels $1$, $\omega$, and $\omega^*$, with $\omega=e^{2\pi i/3}$. 
\label{fig:NbSe2}}
\end{figure*}

To retain the microscopic wavefunction structure, we adopt the six-orbital spinless tight-binding model $h_\text{TB}({\bf k})$ constructed in Ref.~\cite{yu2026NbSe2}, whose explicit form and model parameters are presented in the SM~\cite{supp}. The orbital basis is $|\Phi_{{\bf k}}\rangle=(|d_{z^2} \rangle,|d_{xy}\rangle,|d_{x^2-y^2}\rangle,|p_z^-\rangle,|p_x^+\rangle,|p_y^+\rangle)^T$, where the $d$ orbitals arise from Nb. The $p$ orbitals considered here are combinations between the top and bottom Se atoms, where $p_\alpha^\pm=(p_\alpha^{\mathrm{top}}\pm p_\alpha^{\mathrm{bottom}})/\sqrt{2}$ for $\alpha\in\{x,y,z\}$. We further update $h_\text{TB}({\bf k})$ to a spinful version $H_\text{TB}({\bf k})$ with an Ising-type spin-orbit coupling (SOC) term $\lambda_\text{SOC}L_zS_z$, where the SOC strength $\lambda_\text{SOC}=0.0784$ eV~\cite{liu2013three,he2018magnetic}. Notably, the two spin sectors remain decoupled and are related by TRS. In Fig.~\ref{fig:NbSe2} (a), we present both the lattice structure and the low-energy normal-state band structure of $1H$-NbSe$_2$ from our tight-binding model $H_\text{TB}({\bf k})$. 

We next describe the superconductivity of $1H$-NbSe$_2$ by updating $H_\text{TB}({\bf k})$ to a BdG form with a conventional $s$-wave spin-singlet pairing order. Note that there exist two inequivalent types of local magnetic impurities, which can either sit on a Nb site or a Se site. We therefore introduce the orbital projectors $\Pi_{\rm Nb}=\sigma_+\otimes I_3$ and $\Pi_{\rm Se}=\sigma_-\otimes I_3$, where $\sigma_\pm=(\sigma_0\pm\sigma_z)/2$. Focusing on the spin-up sector as above, the $T$ matrix restricted to either Nb or Se subspace is
\begin{equation}
T_\zeta(E)=\left[J^{-1}\tau_0\otimes I_3-\mathcal{G}_\zeta(E)\right]^{-1},
\end{equation}
where $\zeta\in\{\mathrm{Nb},\mathrm{Se}\}$. In the weak-pairing limit, the impurity-projected Green's function is
\begin{equation}
\mathcal{G}_\zeta(E)\simeq
-\pi\nu_F
\frac{E\tau_0+\Delta\tau_x}{\sqrt{\Delta^2-E^2}}
\otimes Q_{\rm FS}^{\zeta},
\end{equation}
where $Q_{\rm FS}^{\zeta}$ is the $3\times 3$ represnetation of $\Pi_\zeta Q_{\rm FS}\Pi_\zeta$ within the corresponding impurity orbital subspace. Since $\mathrm{Tr}Q_{\rm FS}= \mathrm{Tr}Q_{\rm FS}^\text{Nb} + \mathrm{Tr} Q_{\rm FS}^\text{Se}=1$, $\mathrm{Tr} Q_{\rm FS}^\zeta$ measures the fraction of the target-band Fermi-surface weight accessible to the impurity $\zeta$.
The nonzero eigenvalues of $Q_{\rm FS}^{\zeta}$ thus enter directly into the YSR bound-state energies in Eq.~\ref{eq:BCS_YSR_energy}.

We first evaluate $Q_{\rm FS}^{\rm Nb}$ and $Q_{\rm FS}^{\rm Se}$ directly from the normal-state Fermi surface at $\mu=0$. For either type of impurity, $C_3$ remains a site symmetry, so the eigenstates of $Q_{\rm FS}^{\zeta}$ can be labeled by the orbital $C_3$ eigenvalues \{$1$, $\omega$, $\omega^*$\}, with $\omega=e^{i2\pi/3}$. Notably, $\omega$ and $\omega^*$ channels originate from the $E$ doublet in the vanishing SOC limit, and become weakly split by the Ising SOC. Numerically, we find that
\begin{eqnarray}
\left(\lambda_{1}^{\rm Nb},\lambda_{\omega}^{\rm Nb},\lambda_{\omega^*}^{\rm Nb}\right)
&\simeq (0.3366,\,0.2272,\,0.2216), \nonumber \\
\left(\lambda_{1}^{\rm Se},\lambda_{\omega}^{\rm Se},\lambda_{\omega^*}^{\rm Se}\right)
&\simeq (0.1043,\,0.0542,\,0.0562).
\end{eqnarray}
Since all three eigenvalues are nonzero, our $Q_\text{FS}$ theory thus predicts {\it three YSR pairs for either Nb or Se impurity}. In both cases, the $C_3=1$ channel has the largest $\lambda$ eigenvalue and thus generates the most strongly bound YSR pair, while the YSR pairs from $\omega$ and $\omega^*$ channels are nearly degenerate and less bound. Moreover, each eigenvalue of $Q_{\rm FS}^{\rm Se}$ is substantially smaller than its Nb counterpart, and $\mathrm{Tr}Q_{\rm FS}^{\rm Nb}\gg\mathrm{Tr}Q_{\rm FS}^{\rm Se}$ reflects the much larger total FS weight accessible to a Nb impurity. Therefore, we expect much weaker YSR binding for a Se impurity at the same exchange coupling $J$.

Next, we calculate the YSR spectra by exact diagonalization of the full six-orbital BdG tight-binding Hamiltonian on a $51\times51$ lattice with a Nb- or Se-site magnetic impurity. At $\mu=0$, Fig.~\ref{fig:NbSe2} (b) shows the bound-state spectrum as a function of $J$ for a Nb impurity. Just as expected, three particle-hole-related YSR pairs appear and are numerically classified into the $C_3=1$, $\omega$, and $\omega^*$ channels. The $C_3=1$ pair has the largest binding energy, consistent with its larger $\lambda$ value. The $\omega$ and $\omega^*$ pairs are nearly degenerate with a small SOC-induced splitting of around $10^{-2}\Delta$, as resolved in the inset of Fig.~\ref{fig:NbSe2} (b). As shown in the SM~\cite{supp}, setting $\lambda_{\rm SOC}=0$ restores their twofold degeneracy. For a Se impurity in Fig.~\ref{fig:NbSe2} (c), three YSR pairs are similarly found, but they all remain much closer to the gap edges over the same range of $J$. The full BdG spectra therefore confirm the channel counting, SOC splitting, and pronounced Nb--Se contrast predicted from the eigenspectrum of $Q_{\rm FS}$.

With both $\lambda_a^\zeta$ and the Fermi-level DOS $\nu_F$, we can immediately predict the YSR energies through Eq.~\ref{eq:BCS_YSR_energy} and further examine the accuracy of our quantum geometric theory by comparing with the exact results. Fixing $J=0.30\,\mathrm{eV}$, we calculate the YSR energies using the full six-orbital $T$ matrix by scanning $\mu$ across the entire target band. Figs.~\ref{fig:NbSe2} (d) and (e) compare the binding energy $E_b\equiv\Delta-E_{\rm YSR}$ of the positive-energy member of each symmetry-resolved YSR pair for Nb and Se impurities, respectively. Specifically, the solid curves denote the $Q_{\rm FS}$-based predictions, while the open symbols are obtained from the full $T$-matrix calculation. Remarkably, over most of the $\mu$ range, the two calculations show close quantitative agreement for all three symmetry channels.

Interestingly, Figs.~\ref{fig:NbSe2} (d) and (e) also feature four sharp structures in the $Q_{\rm FS}$-based predictions that deviate from the exact results. We find them to coincide with the Lifshitz transitions of the normal state, as marked by the horizontal dashed lines in Fig.~\ref{fig:NbSe2} (a). In particular, the first two correspond to SOC-split Lifshitz transitions $\mu_S$ and $\mu_{\bar S}$ at generic momenta, while $\mu_M$ is associated with a saddle-point transition near $M$ and $\mu_\Gamma$ with the disappearance of the $\Gamma$-centered Fermi pocket. Near these transitions, the spectral density $\rho(\xi)$ varies rapidly with energy, which thus goes beyond the approximation $\rho(\xi)\simeq\rho(0)$ for Eq.~\ref{eq:BCS_YSR_energy}.

Thus far, we have numerically established that the target band produces three $C_3$-labeled Nb-centered YSR branches, but the 12-band calculation does not make transparent why this particular Nb-derived band realizes such a three-channel structure. Notably, Ref.~\cite{yu2026NbSe2} proposed a downfolded three-orbital Wannier model for $1H$-NbSe$_2$ in the Nb-centered basis $(|\widetilde d_{z^2}\rangle,|\widetilde d_{xy}\rangle,|\widetilde d_{x^2-y^2}\rangle)$, which provides a simple analytic description of the target-band wavefunction structure. Since this model retains only Nb-centered Wannier orbitals as explicit degrees of freedom, it directly captures the Nb-site impurity physics, while Se-site impurities lie beyond this reduced orbital space and require the full six-orbital description. 

As detailed in the SM~\cite{supp}, a unitary transformation of the three $\widetilde d$ orbitals defines three $C_3$-related hybrid orbitals $(|\widetilde c_1\rangle,|\widetilde c_2\rangle,|\widetilde c_3\rangle)$ on which the target-band compact state has equal weight. The threefold rotation permutes these hybrid orbitals as $\widehat C_3|\widetilde c_j\rangle=|\widetilde c_{j-1}\rangle$, with $j$ defined modulo $3$, while in the vanishing-SOC limit a vertical mirror leaves $|\widetilde c_1\rangle$ invariant and exchanges $|\widetilde c_2\rangle$ and $|\widetilde c_3\rangle$. In this basis, the corresponding FS-averaged projector $\widetilde Q_{\rm FS}$ takes the form
\begin{equation}
\widetilde Q_{\rm FS}
=\frac{1}{3}
\begin{pmatrix}
1 & f_F & f_F\\
f_F & 1 & f_F\\
f_F & f_F & 1
\end{pmatrix},
\end{equation}
where $f_F$ denotes the real, DOS-weighted nearest-neighbor phase average over the Fermi surface~\cite{supp}. Diagonalizing $\widetilde Q_{\rm FS}$ yields
\begin{eqnarray}
\lambda_{A_1}
=\frac{1+2f_F}{3}, \qquad
\lambda_E
=\frac{1-f_F}{3},
\end{eqnarray}
where the symmetric state $|\phi_{A_1}\rangle=(1,1,1)^T/\sqrt{3}$ is invariant under both $C_3$ and the mirror and therefore belongs to the $A_1$ channel. Meanwhile, $\lambda_E$ corresponds to a pair of degenerate eigenvectors $|\phi_{\omega}\rangle=(1,\omega,\omega^*)^T/\sqrt{3}$ and $|\phi_{\omega^*}\rangle=|\phi_{\omega}\rangle^*$, which carry conjugate $C_3$ eigenvalues and together form the twofold-degenerate $E$ doublet. Upon restoring Ising SOC, $C_3$ is preserved and the original $E$ doublet splits into the $\omega$ and $\omega^*$ channels. The $A_1$ channel now corresponds to $C_3=1$. Taken together, these results analytically proves that the three YSR branches arise from the wavefunction geometry of a single Nb-derived band.

Our results lead to a concrete experimental prediction for YSR physics in monolayer $1H$-NbSe$_2$, which is ready to be tested by high-resolution scanning tunneling microscopy (STM). YSR spectroscopy has previously been demonstrated in layered $2H$-NbSe$_2$~\cite{menard2015coherent,yang2020observation}, while YSR states of in $1H$-NbSe$_2$ remain largely unexplored in experiments. With a superconducting tip, an energy resolution of $\sim50\,\mu{\rm eV}$ or better can be achieved~\cite{pan1998vacuum,ruby2016SCtip}, offering a practical visibility scale of approximately $0.1\Delta$ for the experimental gap $\Delta\sim0.4$ meV in $1H$-NbSe$_2$. Following Figs.~\ref{fig:NbSe2} (b) and (c), we expect STM measurements to reveal two pairs of in-gap bound-state peaks for a Nb-centered magnetic impurity with an effective $J\gtrsim0.2$ eV: one from the $C_3=1$ channel and the other from the nearly degenerate $\omega$ and $\omega^*$ channels. A Se-centered impurity, however, is expected to show only one resolvable pair of bound-state peaks from the $C_3=1$ channel, and only for a sufficiently strong $J\gtrsim0.5$ eV. The other two $C_3$ channels for the Se impurity are highly likely to remain practically indistinguishable from the superconducting coherence peaks due to their very small binding energies. This characteristic ``{\it 2-peak versus 0/1-peak}'' contrast between Nb- and Se-centered impurities would provide direct experimental evidence for the quantum-geometric contribution to YSR physics in this system.

\section{Conclusions \& Discussions}

In summary, we have established that the YSR physics of conventional $s$-wave SCs is governed by a quantum-geometric quantity of the normal state, namely $Q$ in the flat-band limit and $Q_{\rm FS}$ in the weak-pairing limit. The eigensystem of $Q$ or $Q_{\rm FS}$ directly organizes the YSR physics in almost every aspect, including both bound-state energies and the wavefunctions. A single superconducting band with nontrivial quantum geometry can thus support {\it multiple symmetry-distinct YSR pairs} even for an orbital-independent magnetic impurity. This correspondence is exact in the projected flat-band limit and applies to dispersive bands in the weak-pairing limit. As a proof of concept, we have systematically investigated the YSR states of monolayer $1H$-NbSe$_2$. We find that its isolated low-energy band supports three symmetry-distinct YSR channels, as predicted by our theory. Our framework also paves the way for exploring possible quantum-geometric origins of YSR states in other real-world superconductors.

In this work, we have focused on a classical magnetic impurity that couples identically to all local orbitals at the impurity site. Realistic magnetic impurities, especially transition-metal adatoms, may exhibit orbital-dependent couplings and support multiple scattering channels arising from their internal orbital structure~\cite{ysr2021felix}. This process originates in the microscopic structure of the magnetic impurity itself and is fundamentally different from our quantum-geometry-based mechanism. Nonetheless, understanding the interplay between these two mechanisms for YSR multiplets is an important and intriguing direction for future explorations. Meanwhile, we also expect a natural generalization of our theory to unconventional superconductors, where nonmagnetic scalar impurities can also induce bound states. We leave these interesting questions to future work.  \\

\textit{Note added.}- During the final preparation of this manuscript, we became aware of a recent preprint~\cite{cardoso2026impurity} investigating the relation between quantum geometry and the localization properties of YSR states in flat-band superconductors.

\section{Acknowledgement}

We thank J. Yu and H.H. Weitering for helpful discussions. This work is supported by the U.S. Department of Energy, Office of Science, Office of Basic Energy Sciences, Experimental Condensed Matter Physics Program, under Award No.~DE-SC0026313.

\bibliography{ref}

@article{yu1965bound,
  title={Bound state in superconductors with paramagnetic impurities},
  author={Yu, Luh},
  journal={Acta Physica Sinica},
  volume={21},
  number={1},
  pages={75},
  year={1965},
  doi={10.7498/aps.21.75},
  publisher={Acta Physica Sinica, Chinese Physical Society and Institute of Physics~…}
}

@article{shiba1968,
    author = {Shiba, Hiroyuki},
    title = {Classical Spins in Superconductors},
    journal = {Progress of Theoretical Physics},
    volume = {40},
    number = {3},
    pages = {435-451},
    year = {1968},
    month = {09},
    issn = {0033-068X},
    doi = {10.1143/PTP.40.435},
    url = {https://doi.org/10.1143/PTP.40.435}
}

@article{rusinov1969,
  title={Superconductivity near a paramagnetic impurity},
  author={Rusinov, AI},
  journal={Soviet Journal of Experimental and Theoretical Physics Letters},
  volume={9},
  pages={85},
  url={},
  year={1969}
}

@article{yazdani1997probing,
  title={Probing the local effects of magnetic impurities on superconductivity},
  author={Yazdani, Ali and Jones, BA and Lutz, CP and Crommie, MF and Eigler, DM},
  journal={Science},
  volume={275},
  number={5307},
  pages={1767--1770},
  year={1997},
  url={https://www.science.org/doi/10.1126/science.275.5307.1767},
  publisher={American Association for the Advancement of Science}
}

@article{menard2015coherent,
  title={Coherent long-range magnetic bound states in a superconductor},
  author={M{\'e}nard, Gerbold C and Guissart, S{\'e}bastien and Brun, Christophe and Pons, St{\'e}phane and Stolyarov, Vasily S and Debontridder, Fran{\c{c}}ois and Leclerc, Matthieu V and Janod, Etienne and Cario, Laurent and Roditchev, Dimitri and others},
  journal={Nature Physics},
  volume={11},
  number={12},
  pages={1013--1016},
  year={2015},
  url={https://doi.org/10.1038/nphys3508},
  publisher={Nature Publishing Group UK London}
}

@article{simon2022focusing,
  title = {Quasiparticle focusing of bound states in two-dimensional $s$-wave superconductors},
  author = {Uldemolins, Mateo and Mesaros, Andrej and Simon, Pascal},
  journal = {Phys. Rev. B},
  volume = {105},
  issue = {14},
  pages = {144503},
  numpages = {13},
  year = {2022},
  month = {Apr},
  publisher = {American Physical Society},
  doi = {10.1103/PhysRevB.105.144503},
  url = {https://link.aps.org/doi/10.1103/PhysRevB.105.144503}
}

@article{hoffman2002imaging,
  title={Imaging quasiparticle interference in Bi2Sr2CaCu2O8+ $\delta$},
  author={Hoffman, JE and McElroy, K and Lee, D-H and Lang, KM and Eisaki, H and Uchida, S-I and Davis, JC},
  journal={Science},
  volume={297},
  number={5584},
  pages={1148--1151},
  year={2002},
  url={https://www.science.org/doi/full/10.1126/science.1072640},
  publisher={American Association for the Advancement of Science}
}

@article{mcelroy2003relating,
  title={Relating atomic-scale electronic phenomena to wave-like quasiparticle states in superconducting Bi2Sr2CaCu2O8+ $\delta$},
  author={McElroy, K and Simmonds, RW and Hoffman, JE and Lee, D-H and Orenstein, J and Eisaki, H and Uchida, S and Davis, JC},
  journal={Nature},
  volume={422},
  number={6932},
  pages={592--596},
  year={2003},
  url={https://doi.org/10.1038/nature01496},
  publisher={Nature Publishing Group UK London}
}

@article{hanaguri2010unconventional,
  title={Unconventional s-wave superconductivity in Fe (Se, Te)},
  author={Hanaguri, T and Niitaka, S and Kuroki, K and Takagi, H},
  journal={Science},
  volume={328},
  number={5977},
  pages={474--476},
  year={2010},
  url={https://www.science.org/doi/10.1126/science.1187399},
  publisher={American Association for the Advancement of Science}
}

@article{ming2023evidence,
  title={Evidence for chiral superconductivity on a silicon surface},
  author={Ming, Fangfei and Wu, X and Chen, C and Wang, Kedong D and Mai, Peizhi and Maier, Thomas A and Strockoz, J and Venderbos, JWF and Gonz{\'a}lez, Cesar and Ortega, Jose and others},
  journal={Nature Physics},
  volume={19},
  number={4},
  pages={500--506},
  year={2023},
  url={https://doi.org/10.1038/s41567-022-01889-1},
  publisher={Nature Publishing Group UK London}
}

@article{wu2026prx,
  title = {Microscopic Fingerprint of Chiral Superconductivity},
  author = {Wu, Xuefeng and Hao, Xuan and Chen, Zhuo and Cai, Yuchang and Wu, Minghao and Chen, Congrun and Wang, Kedong and Ming, Fangfei and Johnston, Steven and Zhang, Rui-Xing and Weitering, Hanno H.},
  journal = {Phys. Rev. X},
  volume = {16},
  issue = {1},
  pages = {011026},
  numpages = {14},
  year = {2026},
  month = {Feb},
  publisher = {American Physical Society},
  doi = {10.1103/jmmf-mpr8},
  url = {https://link.aps.org/doi/10.1103/jmmf-mpr8}
}

@article{wang2003qpi,
  title = {Quasiparticle scattering interference in high-temperature superconductors},
  author = {Wang, Qiang-Hua and Lee, Dung-Hai},
  journal = {Phys. Rev. B},
  volume = {67},
  issue = {2},
  pages = {020511(R)},
  numpages = {4},
  year = {2003},
  month = {Jan},
  publisher = {American Physical Society},
  doi = {10.1103/PhysRevB.67.020511},
  url = {https://link.aps.org/doi/10.1103/PhysRevB.67.020511}
}

@article{balatsky2006rmp,
  title = {Impurity-induced states in conventional and unconventional superconductors},
  author = {Balatsky, A. V. and Vekhter, I. and Zhu, Jian-Xin},
  journal = {Rev. Mod. Phys.},
  volume = {78},
  issue = {2},
  pages = {373--433},
  numpages = {0},
  year = {2006},
  month = {May},
  publisher = {American Physical Society},
  doi = {10.1103/RevModPhys.78.373},
  url = {https://link.aps.org/doi/10.1103/RevModPhys.78.373}
}

@article{kreisel2015qpi,
  title = {Interpretation of Scanning Tunneling Quasiparticle Interference and Impurity States in Cuprates},
  author = {Kreisel, A. and Choubey, Peayush and Berlijn, T. and Ku, W. and Andersen, B. M. and Hirschfeld, P. J.},
  journal = {Phys. Rev. Lett.},
  volume = {114},
  issue = {21},
  pages = {217002},
  numpages = {6},
  year = {2015},
  month = {May},
  publisher = {American Physical Society},
  doi = {10.1103/PhysRevLett.114.217002},
  url = {https://link.aps.org/doi/10.1103/PhysRevLett.114.217002}
}

@article{ruby2016orbital,
  title = {Orbital Picture of Yu-Shiba-Rusinov Multiplets},
  author = {Ruby, Michael and Peng, Yang and von Oppen, Felix and Heinrich, Benjamin W. and Franke, Katharina J.},
  journal = {Phys. Rev. Lett.},
  volume = {117},
  issue = {18},
  pages = {186801},
  numpages = {5},
  year = {2016},
  month = {Oct},
  publisher = {American Physical Society},
  doi = {10.1103/PhysRevLett.117.186801},
  url = {https://link.aps.org/doi/10.1103/PhysRevLett.117.186801}
}

@article{kim2020long,
  title={Long-range focusing of magnetic bound states in superconducting lanthanum},
  author={Kim, Howon and R{\'o}zsa, Levente and Schreyer, Dominik and Simon, Eszter and Wiesendanger, Roland},
  journal={Nature communications},
  volume={11},
  number={1},
  pages={4573},
  year={2020},
  url={https://doi.org/10.1038/s41467-020-18406-8},
  publisher={Nature Publishing Group UK London}
}

@article{thouless1982,
  title = {Quantized Hall Conductance in a Two-Dimensional Periodic Potential},
  author = {Thouless, D. J. and Kohmoto, M. and Nightingale, M. P. and den Nijs, M.},
  journal = {Phys. Rev. Lett.},
  volume = {49},
  issue = {6},
  pages = {405--408},
  numpages = {0},
  year = {1982},
  month = {Aug},
  publisher = {American Physical Society},
  doi = {10.1103/PhysRevLett.49.405},
  url = {https://link.aps.org/doi/10.1103/PhysRevLett.49.405}
}

@article{haldane1988,
  title = {Model for a Quantum Hall Effect without Landau Levels: Condensed-Matter Realization of the "Parity Anomaly"},
  author = {Haldane, F. D. M.},
  journal = {Phys. Rev. Lett.},
  volume = {61},
  issue = {18},
  pages = {2015--2018},
  numpages = {0},
  year = {1988},
  month = {Oct},
  publisher = {American Physical Society},
  doi = {10.1103/PhysRevLett.61.2015},
  url = {https://link.aps.org/doi/10.1103/PhysRevLett.61.2015}
}

@article{provost1980riemannian,
  title={Riemannian structure on manifolds of quantum states},
  author={Provost, Jean-Pierre and Vall{\'e}e, Georges},
  journal={Communications in Mathematical Physics},
  volume={76},
  number={3},
  pages={289--301},
  year={1980},
  url={https://doi.org/10.1007/BF02193559},
  publisher={Springer}
}

@article{yu2025quantum,
  title={Quantum geometry in quantum materials},
  author={Yu, Jiabin and Bernevig, B Andrei and Queiroz, Raquel and Rossi, Enrico and T{\"o}rm{\"a}, P{\"a}ivi and Yang, Bohm-Jung},
  journal={npj Quantum Materials},
  volume={10},
  number={1},
  pages={101},
  year={2025},
  url={https://doi.org/10.1038/s41535-025-00801-3},
  publisher={Nature Publishing Group UK London}
}

@article{torma2023essay,
  title = {Essay: Where Can Quantum Geometry Lead Us?},
  author = {T\"orm\"a, P\"aivi},
  journal = {Phys. Rev. Lett.},
  volume = {131},
  issue = {24},
  pages = {240001},
  numpages = {7},
  year = {2023},
  month = {Dec},
  publisher = {American Physical Society},
  doi = {10.1103/PhysRevLett.131.240001},
  url = {https://link.aps.org/doi/10.1103/PhysRevLett.131.240001}
}

@article{liu2025quantum,
  title={Quantum geometry in condensed matter},
  author={Liu, Tianyu and Qiang, Xiao-Bin and Lu, Hai-Zhou and Xie, XC},
  journal={National Science Review},
  volume={12},
  number={3},
  pages={nwae334},
  year={2025},
  url={https://doi.org/10.1093/nsr/nwae334},
  publisher={Oxford University Press}
}

@article{sodemann2015dipole,
  title = {Quantum Nonlinear Hall Effect Induced by Berry Curvature Dipole in Time-Reversal Invariant Materials},
  author = {Sodemann, Inti and Fu, Liang},
  journal = {Phys. Rev. Lett.},
  volume = {115},
  issue = {21},
  pages = {216806},
  numpages = {5},
  year = {2015},
  month = {Nov},
  publisher = {American Physical Society},
  doi = {10.1103/PhysRevLett.115.216806},
  url = {https://link.aps.org/doi/10.1103/PhysRevLett.115.216806}
}

@article{yu2024non,
  title={Non-trivial quantum geometry and the strength of electron--phonon coupling},
  author={Yu, Jiabin and Ciccarino, Christopher J and Bianco, Raffaello and Errea, Ion and Narang, Prineha and Bernevig, B Andrei},
  journal={Nature Physics},
  volume={20},
  number={8},
  pages={1262--1268},
  year={2024},
  url={https://doi.org/10.1038/s41567-024-02486-0},
  publisher={Nature Publishing Group UK London}
}

@article{peotta2015superfluidity,
  title={Superfluidity in topologically nontrivial flat bands},
  author={Peotta, Sebastiano and T{\"o}rm{\"a}, P{\"a}ivi},
  journal={Nature communications},
  volume={6},
  number={1},
  pages={8944},
  year={2015},
  url={https://doi.org/10.1038/ncomms9944},
  publisher={Nature Publishing Group UK London}
}

@article{ugeda2016characterization,
  title={Characterization of collective ground states in single-layer NbSe 2},
  author={Ugeda, Miguel M and Bradley, Aaron J and Zhang, Yi and Onishi, Seita and Chen, Yi and Ruan, Wei and Ojeda-Aristizabal, Claudia and Ryu, Hyejin and Edmonds, Mark T and Tsai, Hsin-Zon and others},
  journal={Nature Physics},
  volume={12},
  number={1},
  pages={92--97},
  year={2016},
  url={https://doi.org/10.1038/nphys3527},
  publisher={Nature Publishing Group UK London}
}

@article{xi2016ising,
  title={Ising pairing in superconducting NbSe 2 atomic layers},
  author={Xi, Xiaoxiang and Wang, Zefang and Zhao, Weiwei and Park, Ju-Hyun and Law, Kam Tuen and Berger, Helmuth and Forr{\'o}, L{\'a}szl{\'o} and Shan, Jie and Mak, Kin Fai},
  journal={Nature Physics},
  volume={12},
  number={2},
  pages={139--143},
  year={2016},
  url={https://doi.org/10.1038/nphys3538},
  publisher={Nature Publishing Group UK London}
}

@article{yu2026NbSe2,
  title = {Quantum geometry in the ${\mathrm{NbSe}}_{2}$ family: Obstructed compact Wannier function and perturbation theory},
  author = {Yu, Jiabin and Jiang, Yi and Xu, Yuanfeng and C\ifmmode \u{a}\else \u{a}\fi{}lug\ifmmode \u{a}\else \u{a}\fi{}ru, Dumitru and Hu, Haoyu and Guo, Haojie and Sajan, Sandra and Wang, Yongsong and Ugeda, Miguel M. and De Juan, Fernando and Bernevig, B. Andrei},
  journal = {Phys. Rev. B},
  volume = {114},
  issue = {12},
  pages = {125104},
  numpages = {13},
  year = {2026},
  month = {Aug},
  publisher = {American Physical Society},
  doi = {10.1103/2bjl-p1vd},
  url = {https://link.aps.org/doi/10.1103/2bjl-p1vd}
}

@article{cualuguaru2026observation,
  title={Observation of an obstructed atomic band in a transition metal dichalcogenide},
  author={C{\u{a}}lug{\u{a}}ru, Dumitru and Jiang, Yi and Guo, Haojie and Sajan, Sandra and Wang, Yongsong and Hu, Haoyu and Yu, Jiabin and Bernevig, B Andrei and de Juan, Fernando and Ugeda, Miguel M},
  journal={Nature Physics},
  pages={1--6},
  year={2026},
  url={https://doi.org/10.1038/s41567-026-03196-5},
  publisher={Nature Publishing Group UK London}
}

@article{holbrook2026real,
  title={Real-space imaging of the band topology of transition metal dichalcogenides},
  author={Holbrook, Madisen and Ingham, Julian and Kaplan, Daniel and Holtzman, Luke N and Bierman, Brenna and Hou, Bowen and Olsen, Nicholas and Nashabeh, Luca and Li, Yiliu and Liu, Song and others},
  journal={Nature Physics},
  pages={1--6},
  year={2026},
  url={https://doi.org/10.1038/s41567-026-03197-4},
  publisher={Nature Publishing Group UK London}
}

@footnote{supp,
note={See the Supplemental Material for detailed information.}
}

@article{zhu2023scattering,
  title = {Scattering theory of delicate topological insulators},
  author = {Zhu, Penghao and Noh, Jiho and Liu, Yingkai and Hughes, Taylor L.},
  journal = {Phys. Rev. B},
  volume = {107},
  issue = {19},
  pages = {195110},
  numpages = {14},
  year = {2023},
  month = {May},
  publisher = {American Physical Society},
  doi = {10.1103/PhysRevB.107.195110},
  url = {https://link.aps.org/doi/10.1103/PhysRevB.107.195110}
}

@article{liu2013three,
  title = {Three-band tight-binding model for monolayers of group-VIB transition metal dichalcogenides},
  author = {Liu, Gui-Bin and Shan, Wen-Yu and Yao, Yugui and Yao, Wang and Xiao, Di},
  journal = {Phys. Rev. B},
  volume = {88},
  issue = {8},
  pages = {085433},
  numpages = {10},
  year = {2013},
  month = {Aug},
  publisher = {American Physical Society},
  doi = {10.1103/PhysRevB.88.085433},
  url = {https://link.aps.org/doi/10.1103/PhysRevB.88.085433}
}

@article{he2018magnetic,
  title={Magnetic field driven nodal topological superconductivity in monolayer transition metal dichalcogenides},
  author={He, Wen-Yu and Zhou, Benjamin T and He, James J and Yuan, Noah FQ and Zhang, Ting and Law, Kam Tuen},
  journal={Communications Physics},
  volume={1},
  number={1},
  pages={40},
  year={2018},
  url={https://doi.org/10.1038/s42005-018-0041-4},
  publisher={Nature Publishing Group UK London}
}

@article{yang2020observation,
  title={Observation of short-range Yu-Shiba-Rusinov states with threefold symmetry in layered superconductor 2H-NbSe2},
  author={Yang, Xing and Yuan, Yuan and Peng, Yang and Minamitani, Emi and Peng, Lang and Xian, Jing-Jing and Zhang, Wen-Hao and Fu, Ying-Shuang},
  journal={Nanoscale},
  volume={12},
  number={15},
  pages={8174--8179},
  year={2020},
  url={https://doi.org/10.1039/d0nr01383h},
  publisher={The Royal Society of Chemistry}
}

@article{ruby2016SCtip,
  title = {Experimental Demonstration of a Two-Band Superconducting State for Lead Using Scanning Tunneling Spectroscopy},
  author = {Ruby, Michael and Heinrich, Benjamin W. and Pascual, Jose I. and Franke, Katharina J.},
  journal = {Phys. Rev. Lett.},
  volume = {114},
  issue = {15},
  pages = {157001},
  numpages = {5},
  year = {2015},
  month = {Apr},
  publisher = {American Physical Society},
  doi = {10.1103/PhysRevLett.114.157001},
  url = {https://link.aps.org/doi/10.1103/PhysRevLett.114.157001}
}

@article{pan1998vacuum,
  title={Vacuum tunneling of superconducting quasiparticles from atomically sharp scanning tunneling microscope tips},
  author={Pan, SH and Hudson, EW and Davis, JC},
  journal={Applied Physics Letters},
  volume={73},
  number={20},
  pages={2992--2994},
  year={1998},
  url={https://doi.org/10.1063/1.122654},
  publisher={American Institute of Physics}
}

@article{ysr2021felix,
  title = {Yu-Shiba-Rusinov states in real metals},
  author = {von Oppen, Felix and Franke, Katharina J.},
  journal = {Phys. Rev. B},
  volume = {103},
  issue = {20},
  pages = {205424},
  numpages = {21},
  year = {2021},
  month = {May},
  publisher = {American Physical Society},
  doi = {10.1103/PhysRevB.103.205424},
  url = {https://link.aps.org/doi/10.1103/PhysRevB.103.205424}
}

@article{cardoso2026impurity,
  title={Impurity-dependent quantum geometry in 2D flat band superconductors},
  author={Cardoso, Sim{\~a}o S and Mesaros, A and Simon, P},
  journal={arXiv preprint arXiv:2609.00338},
  url={https://doi.org/10.48550/arXiv.2609.00338},
  year={2026}
}

\end{document}